\documentclass[]{aa}

\usepackage{xcolor}
\usepackage{booktabs}
\usepackage{overpic}
\usepackage{graphicx}
\usepackage{adjustbox}
\usepackage{txfonts}
\usepackage{hyperref}
\hypersetup{
    colorlinks=true,
    linkcolor=blue,
    filecolor=magenta,      
    urlcolor=blue,
    citecolor=blue,
    }
\begin{document} 
\title{Exploring the formation environment and dynamics of multiple stellar populations in Globular Clusters through binary systems.}
\subtitle{}

\author{E.\,Bortolan\inst{1}, J. Bruce\inst{2}, A. P. Milone\inst{1,3}, E. Vesperini\inst{2}, E. Dondoglio\inst{3}, M. V. Legnardi\inst{1}, F. Muratore\inst{1}, T. Ziliotto\inst{1}, G. Cordoni\inst{4},  E. P. Lagioia\inst{5}, A. F. Marino\inst{3}, M. Tailo\inst{3}
}

\institute{Dipartimento di Fisica e Astronomia ``Galileo Galilei'', Univ. di Padova, Vicolo dell'Osservatorio 3, Padova, IT-35122
\and 
Department of Astronomy, Indiana University, Swain West, 727 E. 3rd Street, Bloomington, IN 47405, USA
\and
Istituto Nazionale di Astrofisica - Osservatorio Astronomico di Padova, Vicolo dell’Osservatorio 5, Padova, IT-35122
\and
Research School of Astronomy and Astrophysics, Australian National University, Canberra, ACT 2611, Australia
\and
South-Western Institute for Astronomy Research, Yunnan University, Kunming 650500, PR China
}

\titlerunning{Exploring the formation environment of multiple stellar populations in Globular Clusters through binary systems.} 
\authorrunning{Bortolan et al.}

\date{Received XXXX; accepted XXXX}

\abstract{}{
Globular Clusters (GCs) are known to host distinct stellar populations,  characterized by different chemical compositions. Despite extensive research, the origin of these populations remains elusive. According to many formation scenarios, the second population (2P) originated within a compact and denser region embedded in a more extended first population (1P) system. As a result, 2P binaries should be disrupted at a larger rate than 1P binaries. For this reason, binary systems offer valuable insight into the environments in which these stellar populations formed and evolved.} 
{In this research, we analyze the fraction of binaries among 1P and 2P M dwarfs in the outer region of NGC 288 using Hubble Space Telescope data. We combine our results with those from a previous work, where we inferred the fraction of 1P and 2P binaries in the cluster center.}
{In the outer region, we find a predominance of 1P binaries ($97^{+1}_{-3} \%$) compared to 2P binaries ($3\pm1\%$) corresponding to an incidence of binaries with a mass ratio (i.e., the ratio between the masses of the primary and secondary star) greater than 0.5 equal to $6.4\pm1.7\%$ for the 1P population and $0.3\pm0.2\%$ for the 2P population. 
These binary fractions and incidences differ from those found in the cluster’s central region, where the 1P and 2P populations exhibit similar binary incidences and fractions.

These results are in general agreement with the predictions of simulations following the evolution of binary stars in multiple-population GCs, starting with a dense 2P subsystem concentrated in the central regions of a 1P system.}
{}

\maketitle
%
\section{Introduction}
\label{sec:intro}
Globular Clusters (GCs) were once considered the best approximations of simple stellar population \citep[e.g.\,][]{renzini1986a}. However, various studies have revealed groups of GC stars with different chemical compositions. 
It is now widely accepted that most GCs exhibit multiple stellar sequences and complex horizontal-branch morphology in the color-magnitude diagrams (CMDs) built with appropriate photometric bands. 
All these features are nothing but different facets of the same phenomenon, known as the multiple stellar population phenomenon, according to which GCs host two main stellar populations: a first population (1P) with a Galactic-field-like chemical composition and a second population (2P) enhanced in He, N, Na, and depleted in C and O \citep[see reviews by][]{kraft1994a, gratton2012a, bastian2018a, milone2022a}.

The formation of multiple stellar populations is a topic of intense debate.
According to most 
 scenarios, 2P stars formed in the central regions of the primordial GCs, whereas 1P stars had a more sparse initial spatial distribution \citep[see e.g.][]{dercole08,bekki10,bekki11,renzini2015a,calura19,mkenzi21,lacchin22}. 
In this context, binary stars may be valuable tools to constrain 
 spatial properties of 1P and 2P stars at the time of their formation and during their subsequent dynamical evolution.
For instance, in high-density environments, frequent stellar encounters lead to a high disruption rate of binary systems. Conversely, in low-density environments, the binary fraction is expected to be higher. Therefore, the incidence of binaries among 1P and 2P stars can provide insight into the properties of the environments where these stellar populations formed \citep[e.g.][]{vesperini11, hong15, hong2016a,sollima22,hypki22}.

\begin{table*}[t!]
\caption{Details about data used in this analysis.}
\centering
\begin{tabular}{ccccccc}

\toprule\toprule
\textbf{Filter} & \textbf{Instrument} & \textbf{N $\times$ Exp. time}&\textbf{Radial distance [arcmin]} & \textbf{Program ID}  \\ 
\midrule
F606W & ACS/WFC & $15\ \rm s+3\times200\  \rm s$ &5.99&  12193  \\ 
F814W & ACS/WFC & $10\ \rm s+3\times150\  \rm s$ & 5.99&12193  \\ 
F110W & WFC3/IR & $3\times\ 142 \rm s+5\times1202\  \rm s$ & 5.45& 16289  \\ 
F160W & WFC3/IR & $4\times142\ \rm s+2\times1202\  \rm s+7\times1302\ \rm s$ &5.45& 16289  \\ 
\bottomrule
\end{tabular}

\label{data}
\end{table*}

Early determinations of the incidence of binaries among multiple populations are based on spectroscopy. 
\citet{lucatello2015a} analyzed 21 binaries located in 10 GCs around the half-light radius ($R_{hl}$) and determined that the fraction of 1P binaries is $4.1\ \pm\ 1.7$ times higher than that of 2P binaries. Based on 12 binaries in NGC\,6362 located beyond  $0.5\ R_{hl}$,  \citet{dalessandro2018a} discovered that their fraction among 1P and 2P stars is $4.7\ \pm 1.4\ \%$ and $0.7\ \pm\ 0.7\ \%$, respectively. Similar conclusions are obtained by \cite{dorazi2010a}, who found that four out of the five studied barium stars of NGC\,6121 are sodium poor and oxygen rich. 


 Conversely, \citet{milone2020} investigated the innermost regions in five GCs, namely NGC\,288, NGC\,6121, NGC\,6352, NGC\,6362, and NGC\,6838 from $ 0.3\ R_{hl}$ to about $0.8\ R_{hl}$. 
 Except for NGC 6121, they found that the inner regions are characterized by similar 1P and 2P binary fractions.
  Observations also provided the first evidence of the existence of mixed binaries composed of one 1P star and one 2P star, a population of binaries predicted to form during exchange encounters in binary-single, binary-binary interactions in the study of \cite{hong15, hong2016a}.

In this paper, we analyze the binary systems among the multiple populations of NGC\,288, a GC that has been extensively studied in the context of multiple stellar populations. Previous researches based on both spectroscopy and photometry have identified two groups of stars with different abundances of certain light elements, including helium, carbon, nitrogen, oxygen, sodium, and aluminum \citep[e.g.,][]{shetrone2000a, kayser2008a, smith2009a, carretta2009a, milone2018a, marino2019a}. Additionally, photometric studies have revealed two distinct sequences in the CMD that can be traced continuously across various evolutionary stages, including the red-giant branch, sub-giant branch, and main sequence (MS) \citep[][]{lee2009a, piotto2013a, monelli2013a, milone2017, lagioia2021a, Jang2022, dondoglio}.

All previous studies on binaries have analyzed multiple systems within $1\ R_{hl}$ and with a primary star mass ($M^{1}_{bin}$) more massive than $\sim 0.6\ M_{\odot}$. In the present work, we investigate for the first time binaries in the lower part of the MS, with $M^{1}_{bin}\leq\ 0.5\ M_{\odot} $,  located in the outer regions of NGC 288, one of the five GCs examined by \cite{milone2020}. The outer regions of a GC are particularly interesting because those are where simulations of the dynamical evolution of binaries in multiple-population GCs predict larger differences between the local 1P and 2P binary incidences.

\section{Observations and data reduction}

We analyzed data collected using the F606W and F814W filters of the Wide Field Channel of the Advance Camera for Surveys (ACS/WFC) and F110W and F160W filters of the IR channel of the Wide Field Camera 3 (WFC3/IR) on board the Hubble Space Telescope (HST). The ACS/WFC images are part of the program GO 12193 (PI Lee), while the WFC3/IR images are from the program GO 16289 (PI Milone). Details about these observations are summarized in Table \ref{data}.

For data reduction, we performed effective Point Spread Function photometry following the procedure developed by Jay Anderson \citep{anderson2008} using the KS2 program. In summary, the process begins by detecting the brightest and most isolated stars in the image. Their fluxes and positions are then determined, after which they are subtracted from the image. Subsequent iterations continue to identify, measure, and subtract progressively fainter stars closer to their neighbors. Three methods are available to achieve optimal photometry and astrometry for different stellar luminosity ranges: method I for relatively bright stars, method II for faint stars, and method III for crowded regions. We adopted method I, which provides flux and position for all stars producing distinct peaks within a $5 \times 5$ pixel region, using the best Point Spread Function model corresponding to that position. This procedure is repeated for each exposure, and in the end stellar fluxes and positions are properly averaged together. Stellar magnitudes are calibrated to the Vega mag system by computing the aperture correction to the instrumental magnitudes and applying the photometric zero-points available at the space telescope science institute web page \citep[for more details, see][]{milone23a}. Stellar coordinates are corrected for geometric distortion by using the solutions provided by \cite{anderson2008} and \cite{anderson2022a}. We compared the  positions of stars collected at different epochs to derive the proper motions, which we used to identify the probable cluster members and field stars \citep[see e.g.][for details]{anderson2003a, milone23a}.

\begin{figure*}[t]
    \centering
    \includegraphics[height=.4\textwidth,trim={0cm 0.0cm 0.0cm 0cm},clip]{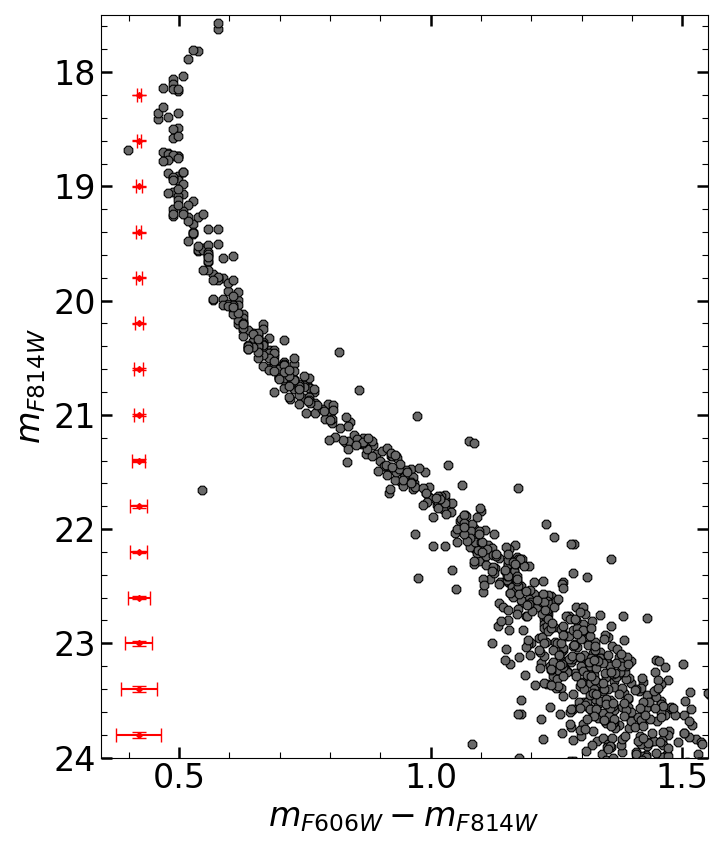}
    \includegraphics[height=.4\textwidth,trim={0cm 0.0cm 0.0cm 0cm},clip]{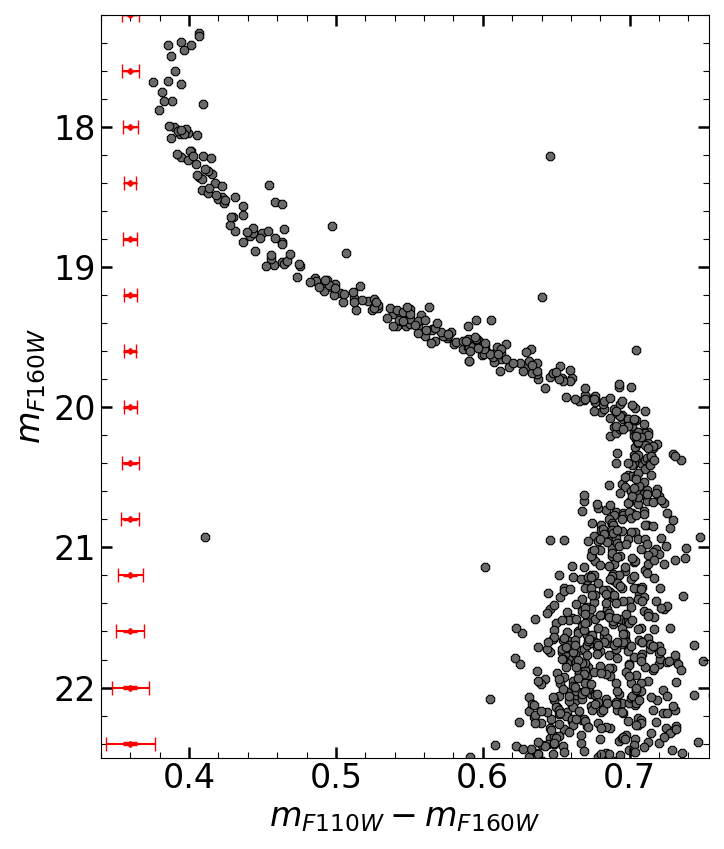}
     \caption{$m_{\rm F814W}$ versus $m_{\rm F606W}-m_{\rm F814W}$ (left) and $m_{\rm F160W}$ versus $m_{\rm F110W}-m_{\rm F160W}$ CMD (right) of NGC\,288 stars. The error bars plotted on the left side of each CMD indicate the typical uncertainties for stars with different magnitudes.
    }
    \label{fig:Fig1}
\end{figure*}

The CMDs $m_{\rm F814W}$ vs. $m_{\rm F606W}-m_{F814W}$ and $m_{\rm F160W}$ vs. $m_{\rm F110W} - m_{\rm F160W}$, derived with this procedure, are shown in the left and right panel in Fig. \ref{fig:Fig1}, respectively.

\label{sec:data}

To estimate the fraction of 1P and 2P binaries, we compared the observed CMD with simulated CMD constructed using artificial stars (AS). The AS test follows the method developed by \cite{anderson2008}. In summary, we generated a sample of AS equal to 10 times the number of observed stars within the instrumental magnitude range of $-13.4$ to $-4.4$ in the F814W filter. The F606W, F110W, and F160W magnitudes were calculated using the colors of the fiducial lines of 1P and 2P MS stars. We reduced these ASs using exactly the same procedure as the one adopted for real stars.
\label{sec:AS}
\begin{figure*}[h!]
    \centering
    \includegraphics[height=.38\textwidth,trim={0cm 0.0cm 0.0cm 0cm},clip]{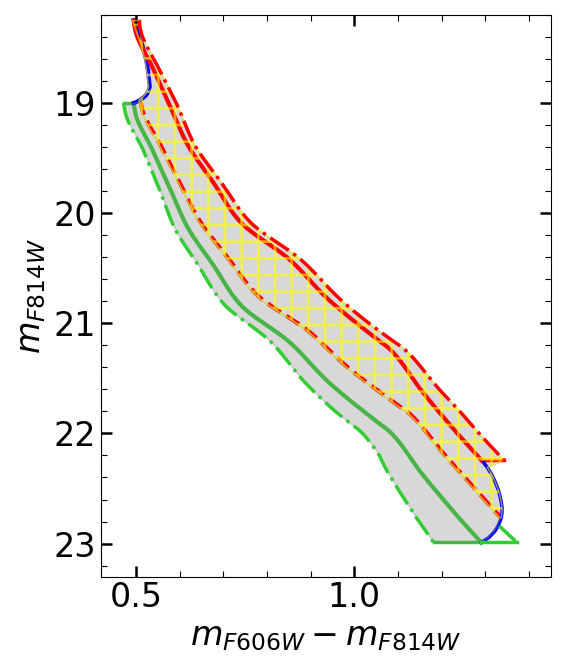}
    \includegraphics[height=.38\textwidth,trim={0cm 0.0cm 0.0cm 0cm},clip]{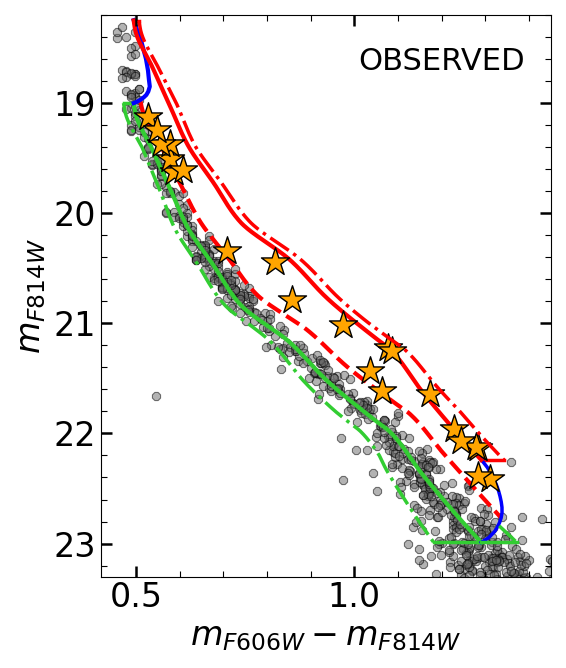}
    \includegraphics[height=.38\textwidth,trim={0cm 0.0cm 0.0cm 0cm},clip]{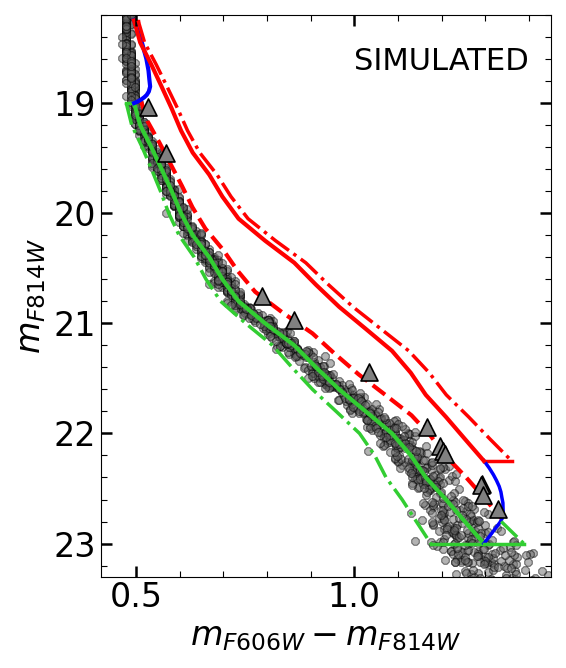}
    
    %
    %
     \caption{
     Procedure to estimate the fraction of binaries among upper MS stars with masses of $0.36-0.75\ M_{\odot}$. 
     This figure reproduces the $m_{\rm F814W}$ versus $m_{\rm F606W}-m_{\rm F814W}$ CMD of Fig.\,\ref{fig:Fig1}.
     The candidate binaries are represented with yellow starred symbols, whereas the remaining stars are plotted with small gray circles.
     The green solid line represents the MS fiducial line, while solid and dashed red lines are the fiducial lines of binaries with mass ratios $q=0.5$ and $q=1$, respectively. The green and red dashed dotted lines are the MS fiducial line and the fiducial line of binaries with $q=1$ shifted by three times the color error. The blue lines mark the locus of binaries where the mass of the primary component corresponds to 0.36 (upper line) and 0.75 $M_{\odot}$ (bottom line) and the mass ratio ranges from 0 to 1. In the simulated CMD (right panel), stars located within the region B are displayed with gray triangles. }
    \label{fig:Fig2}

\end{figure*}\label{sec:3}

\section{Binaries along the main sequence}
Due to the considerable distance of GCs (e.g., $\sim9$ kpc for NGC 288, \citealt{baumgardt2021a}), the {\it HST} is unable to resolve individual components of the binary systems in  these stellar clusters. Consequently,  binaries appear as single-point sources. The magnitude of an unresolved binary system, $m_{bin}$, is determined by the combined fluxes of its components
\[
m_{bin}=m_{1}-2.5\log \left( 1+\frac{F_{2}}{F_{1}} \right)
\]
 with $m_{1}$ the magnitude of the primary star (i.e., the most massive one), and $F_{1}$ and $F_{2}$ the fluxes of the primary and secondary component, respectively.

The procedure used to measure the fraction of binaries with large mass ratio in NGC\,288 is illustrated in the left panel of Fig.\,\ref{fig:Fig2} and is based on the method by \cite{milone2012}. This method relies on the fact that in the $m_{\rm F814W}$ vs.\, $m_{\rm F606W}-m_{\rm F814W}$ CMD of a GC, binary systems composed of two MS stars are located on the red and bright side of the MS, well separated from the bulk of single MS stars. The exact location of the binaries depends on the mass of the primary star and the mass ratio, $q=M_{2}/M_{1}$, with $M_{1}$ and $M_{2}$ being the mass of the primary and secondary star, respectively \citep[see][and references therein]{milone2012, mohandasan2024a}.

As illustrated in the left panel of Fig.\,\ref{fig:Fig2}, we identified two regions in the CMD. Region A (grey shaded area in Fig.\,\ref{fig:Fig2}) contains all single MS stars with $19.0 < m_{F814W} < 23.1$ mag, as well as binary systems where the primary star lies within the same luminosity interval. We defined the left boundary (dash-dotted green line) by shifting the MS fiducial line (solid green line) four times the average color error to the left. The right boundary (dash-dotted red line) corresponds to the equal-mass binaries fiducial (solid red line) shifted by four times the color error to the right.
 The faint and bright boundaries of the region A (blue lines) correspond to the loci of binary systems with $m_{F814W}=19.0$ and $23.0$ mag and \emph{q} ranging from zero to one.
 Region B (yellow dashed area in left panel of Fig.\,\ref{fig:Fig2}) is the portion of the region A that includes the binaries with $q>0.5$. It is limited on the left side of the CMD by the fiducial of binary systems with $q=0.5$ (red dashed line).  

The observed stars are shown in the central  panel of Fig.\,\ref{fig:Fig2}, where we overlap to the $m_{F814W}$ vs $m_{F606W}-m_{F814W}$ CMD the lines defined in the left panel. We used yellow starred symbols to mark the probable binaries in the region B of the CMD.
The simulated CMD for single MS stars that we obtained from AS stars is illustrated in the right panel of Fig.\,\ref{fig:Fig2}.

The fraction of binaries along the MS with $q>0.5$ is calculated as
\begin{equation}
f_{\rm bin}^{q>0.5}=\frac{N_{\rm obs}^{B}}{N_{\rm obs}^{A}}-\frac{N_{\rm sim}^{B}}{N_{\rm sim}^{A}}
\label{eq:1}
\end{equation}
where $N_{obs}$ and $N_{sim}$ represent the number of observed cluster members and simulated stars, respectively.
The resulting binary fraction is $f_{\rm bin}^{MS}=4.51\pm0.99\%$.
 The uncertainty is derived by means of standard error propagation by associating Poisson errors to the star counts.

\begin{figure*}[!h]
    \centering
    \includegraphics[height=.32\textwidth,trim={0cm 0.0cm 0.0cm 0cm},clip]{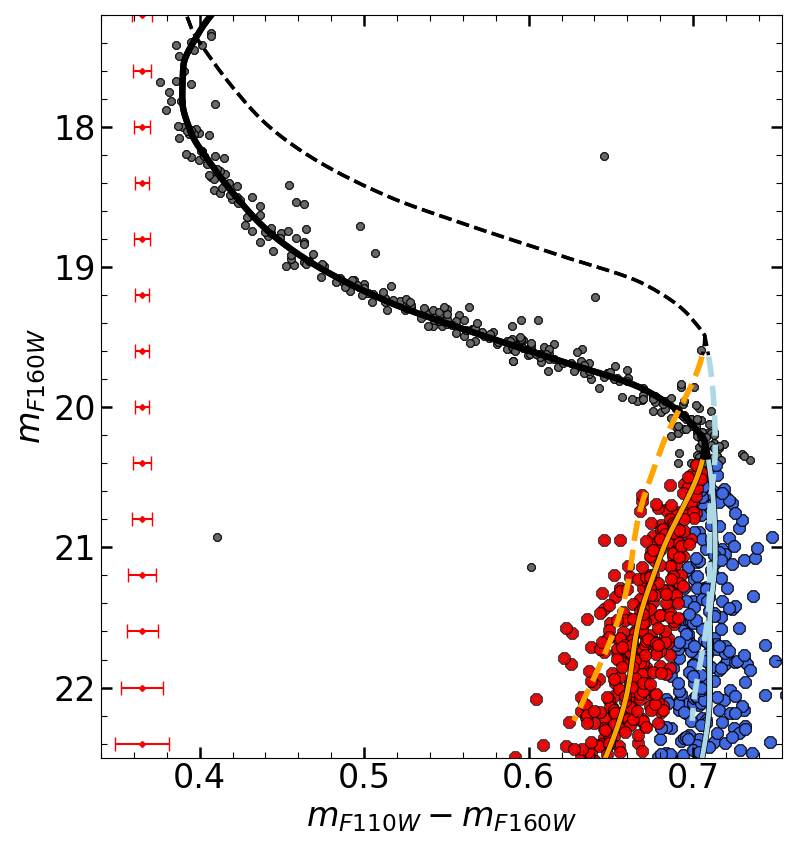  }
    \includegraphics[height=.32\textwidth,trim={0cm 0.0cm 0.0cm 0cm},clip]{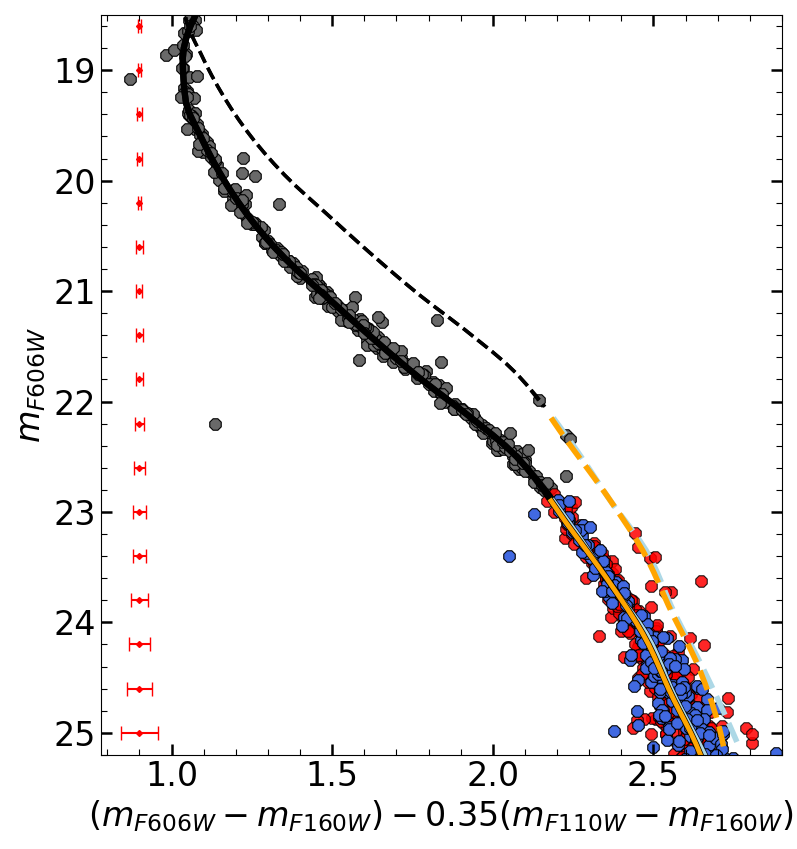 }
     \caption{
     $m_{\rm F160W}$ versus $m_{F110W}-m_{\rm F160W}$ CMD (left) and $m_{\rm F606W}$ versus  $(m_{\rm F606W}-m_{\rm F160W})-0.35\ (m_{\rm F110W}-m_{\rm F160W})$ pseudo-CMD of NGC\,288 stars.
     The probable 1G and 2G M dwarfs, selected from the CMD on the left panel, and the corresponding fiducial lines are colored red and blue, respectively. Dotted lines represent the fiducials of equal-mass binaries (i.e. the MS shifted upward by $0.752$ mag). The remaining stars are colored grey, whereas the fiducial lines of the upper MS are represented by black lines.
     Typical error bars are shown in red on the left.
    }
    \label{fig:Fig3}
\end{figure*}\label{sec:data}

\begin{figure*}[!h]
    \centering
    \includegraphics[width=.3\textwidth,clip]{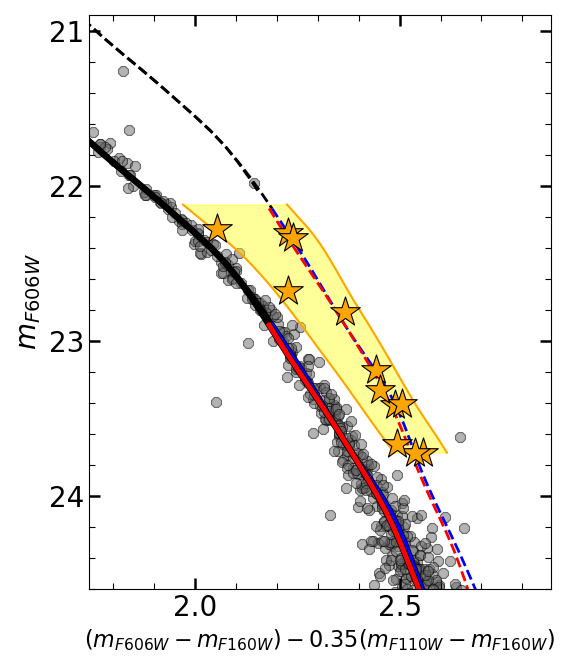}
    \includegraphics[width=.3\textwidth,clip]{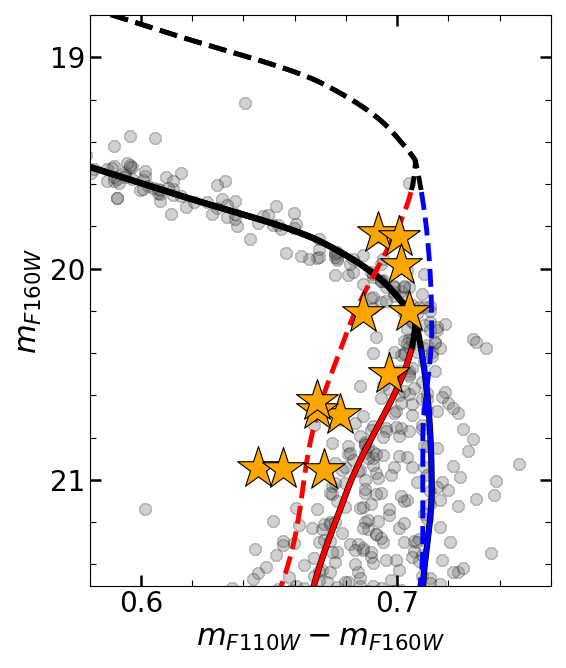}
    \includegraphics[width=.3\textwidth,clip]{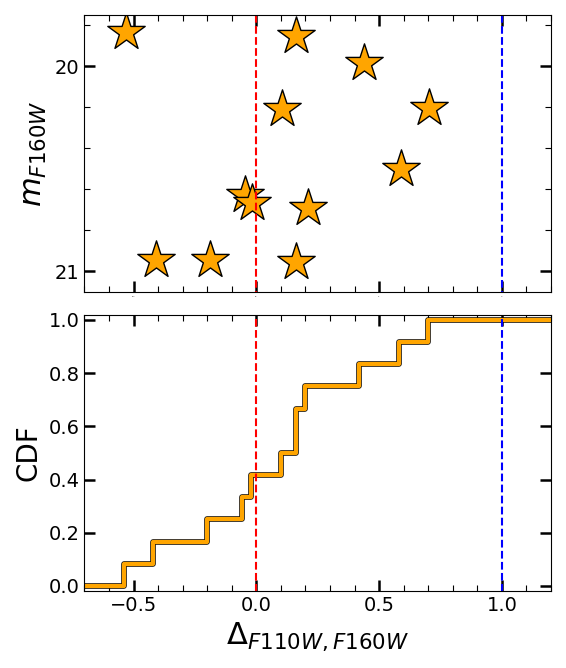}
    \caption{Procedure to select and characterize the binaries used to infer the incidence of 1P and 2P binaries. The left panel is a zoom of the $m_{F606W}$ vs.\,($m_{F606W}-m_{F160W}$)-0.35($m_{F110W}-m_{F160W}$) diagram at the bottom of the MS, whereas the middle panel shows the $m_{F160W}$ vs.\,$m_{F110W}-m_{F160W}$ CMD. The right panels show the verticalized $m_{F160W}$ vs.\,$\Delta_{F110W,F160W}$ diagram (top) and the $\Delta_{F110W,F160W}$ cumulative distribution (bottom) for the probable binaries. The probable binaries, which are represented with starred symbols, are selected from the left-panel diagram, where they lie within the yellow-shaded area of the CMD. The red and blue solid lines are the fiducials of 1P and 2P stars, while the dotted mark the corresponding loci of equal-mass binaries. See the text for details. }
    \label{fig:Fig5}
\end{figure*}
%
%

\section{Binaries among multiple populations }

To estimate the binary fraction among multiple population, we used the method by \cite{milone2020}, which relies on two photometric diagrams, the first allowing the identification of 1P and 2P stars, and the second where the 1P and 2P sequences are nearly superimposed on each other in order to select binaries.

 As first photometric diagram, we used the $m_{\rm F160W}$ versus $m_{\rm F110W} - m_{\rm F160W}$ CMD, which has proven effective in distinguishing the multiple populations among M dwarfs \citep{milone2014, dondoglio}. This is because the F160W luminosity of M dwarfs is significantly affected by flux absorption due to molecules containing oxygen atoms, whereas the F110W flux is minimally dependent by the stellar oxygen abundance.
 As a consequence, the 1P stars, which have higher O abundance than 2P stars, exhibit fainter F160W magnitudes and bluer F110W$-$F160W colors than 2P stars with the same luminosity \citep[e.g.\,][]{milone2012, dotter2015}

\begin{table*}[h!]

\caption{Fraction of binaries with $q>0.5$. The mass range $0.36-0.75\ M_{\odot}$ corresponds to the magnitude range $19.0\leq m_{F814W}\leq23.1$, while the mass range $0.39-0.75\ M_{\odot}$ corresponds to $19.5\leq m_{F606W}\leq24.5$. }
\label{tab:binary fraction}
\centering
\begin{tabular}{ccccc}
\toprule\toprule
& \textbf{$0.36-0.75\ M_{\odot}$} & \textbf{ $0.39-0.75\ M_{\odot}$} \\ 
\midrule
$f^{q>0.5}_{bin}$ & $4.51\pm0.99\%$ & $3.93\pm0.95\%$   \\[3pt]
\bottomrule
\end{tabular}

\end{table*}

\begin{figure*}[!h]
    \centering
    \begin{overpic}[width=.3\textwidth,clip]{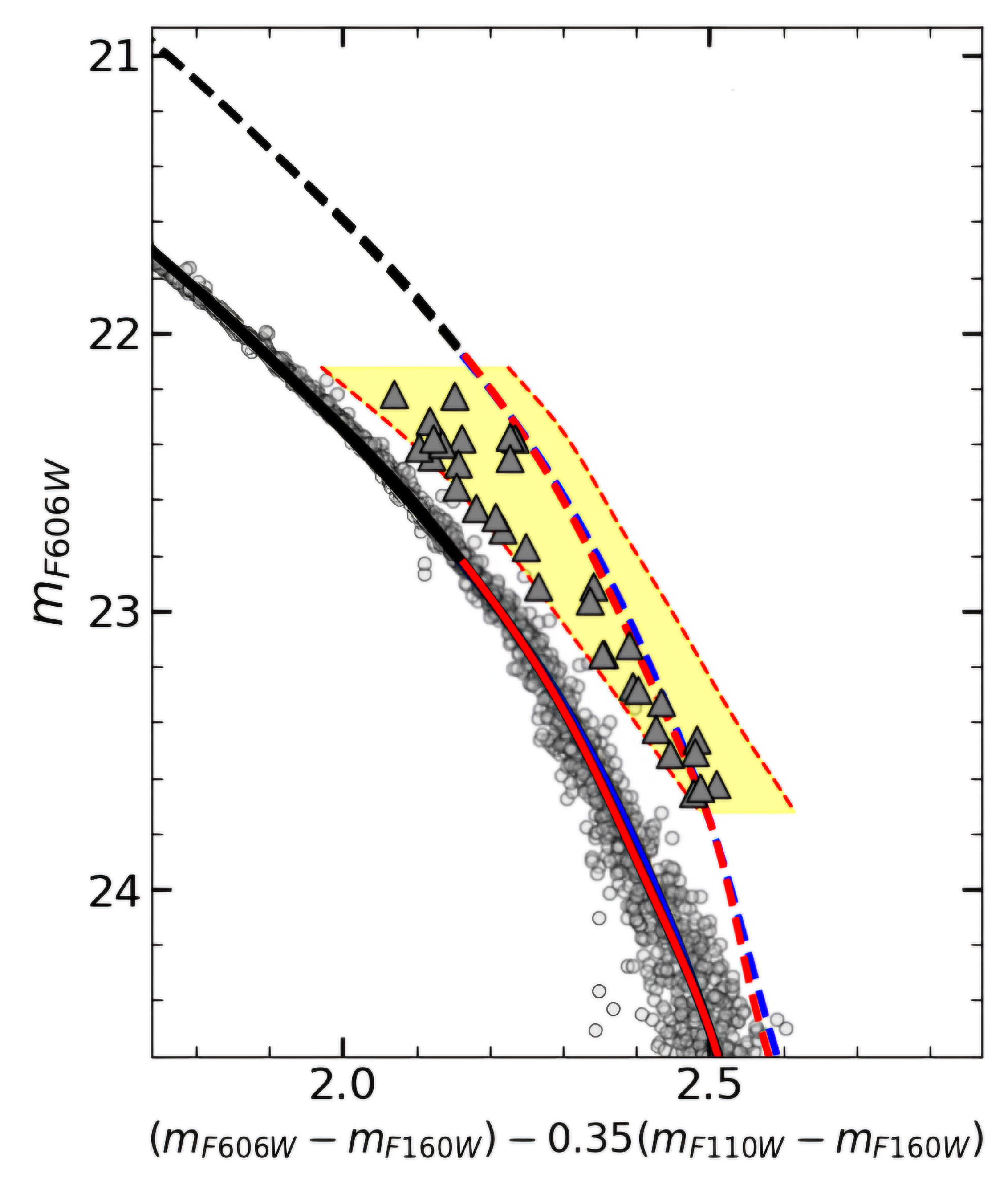}
     \put(48,90){Best-fit model:}
     \put(48,84){$f^{1P}=0.97$ }
     \put(48,79){$f^{2P}=0.03$}
     \put(48,74){$f^{\rm mix}=0.00$}
    \end{overpic}
    \includegraphics[width=.3\textwidth,clip]{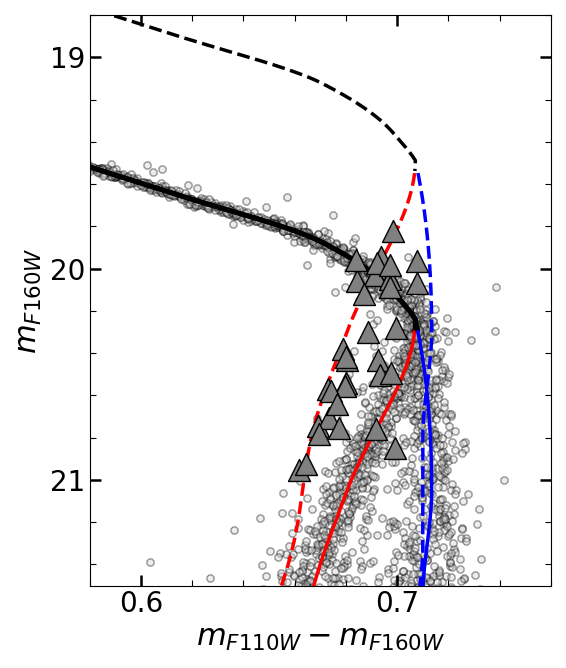 }
    \includegraphics[width=.3\textwidth,clip]{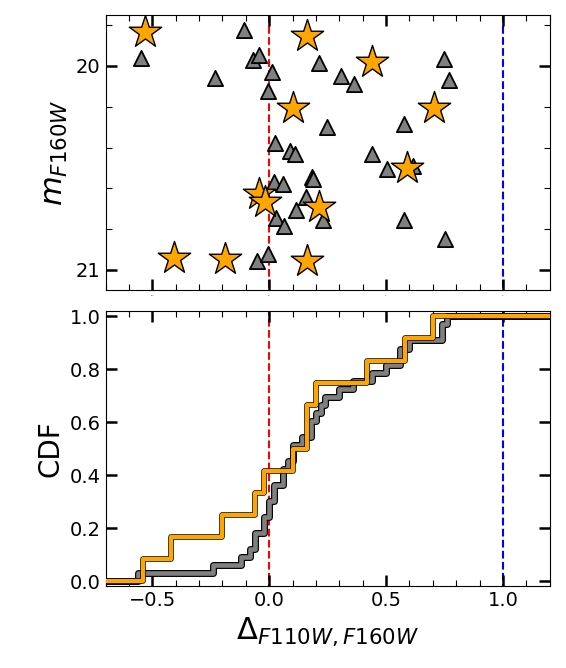}
    \caption{
    Simulated diagrams, following the same layout as Fig.\,\ref{fig:Fig4}, assuming the best-fit values for the fractions of 1P, 2P and mixed binaries as indicated in the left plot. Grey triangles represent artificial binaries used for comparison with binaries selected in Fig.\,\ref{fig:Fig5}. In the right panels, we show both the selected observed binaries, displayed as in Fig.\,\ref{fig:Fig5}, and the artificial binaries, for comparison.
    }
     \label{fig:Fig6}
\end{figure*}

The presence of multiple populations in NGC\,288 is visible in the CMD on the right panel of Fig.\,\ref{fig:Fig1}. Indeed, the color broadening of stars fainter than the MS knee ($m_{F160W}\gtrsim 20.1 $ mag) exceeds what can be explained by photometric errors alone, suggesting the presence of stars with different O abundances \citep[see also][]{dondoglio}.

In Fig.\,\ref{fig:Fig3}, 1P stars are highlighted in red and 2P stars in blue.
Their fiducials are represented by solid orange and cyan lines, respectively, while the corresponding fiducial lines for equal-mass binaries are shown as dashed lines. The fraction of 1P and 2P stars correspond to $59.5\pm3.8\% $ and $40.5\pm2.9\%$, respectively.

As second photometric diagram, used to select the binaries among multiple populations, the we used the $m_{\rm F606W}$ versus $(m_{\rm F606W}-m_{\rm F160W})-0.35\ (m_{\rm F110W}-m_{\rm F160W})$ pseudo-CMD, in which the fiducial lines of 1P adn 2P stars overlap (right panel in Fig.\,\ref{fig:Fig3}).
In this diagram, we selected a sample of 12 candidate binaries that we used to determine the incidence of 1P and 2P binaries. These stars  are located 
 within the dashed yellow area shown in the left panel of Fig.\,\ref{fig:Fig5} and are represented with starred symbols. 
The left boundary of this region is defined by the fiducial line of 2P single MS stars, shifted by four times the photometric error to exclude single MS stars with large errors. The right boundary is defined by the fiducial of 2P equal-mass binaries, shifted by four times the colour error to account for binaries appearing redder due to photometric errors. We set the upper limit to $m_{F606W}=22.1$ mag because it is not possible to distinguish 1P and 2P stars at brighter magnitudes. 
 The lower limit is set to $m_{F606W}=23.75$ mag because, at fainter magnitudes, it is challenging to disentangle binaries and single stars due to the large observational errors.

In the middle panel of Fig.\,\ref{fig:Fig5}, we displayed the selected binaries in the $m_{\rm F160W}$ vs.\,$m_{\rm F110W}-m_{\rm F160W}$ CMD, which is used to derive the verticalized diagram $m_{\rm F160W}$ vs. $\Delta_{\rm F110W, F160W}$. The absissa of this diagram is calculated in such a way that the two fiducials of equal-mass binaries translate into vertical lines. Specifically, we used the relation 
\begin{equation}
\Delta_{\rm F110W, F160W}=(X-X^{\rm 1P-1P})/(X^{\rm 2P-2P}-X^{\rm 1P-1P})
\label{eq:2}
\end{equation}

where $X$ is the F110W$-$F160W colour of the selected binary, and $X^{\rm 1P-1P}$ and $X^{\rm 2P-2P}$ are the colours of the fiducial of equal-mass 1P and 2P binaries, respectively, at the same magnitude as the selected binary. 
The resulting verticalized diagram is shown in the upper right panel of Fig.\,\ref{fig:Fig5} and is used to derive the cumulative distribution function (CDF) of the selected binaries  (lower-right panel of Fig.\,\ref{fig:Fig5}). 
\begin{table*}[h!]
\centering
\caption{Binary fraction derived from the $m_{\rm F606W}$ versus  $(m_{\rm F606W}-m_{\rm F160W})-0.35\ (m_{\rm F110W}-m_{\rm F160W})$ pseudo-CMD for three mass intervals. These  intervals correspond to the three magnitude bins into which the F606W magnitude range used to derive the overall binary fraction was divided (see section \ref{radial} for the details).}

\begin{tabular}{ccccc}
\toprule\toprule
& \textbf{$0.39-0.52\ M_{\odot}$ } & \textbf{$0.52-0.63\ M_{\odot}$}& \textbf{$0.63-0.75\ M_{\odot}$}  \\ 
\midrule
$f^{q>0.5}_{bin}$ & $4.18\pm1.43\%$ & $3.31\pm1.37\%$ & $4.71\pm2.13\%$ &  \\[3pt]

\bottomrule
\end{tabular}

\label{tab:binary fraction2}
\end{table*}

\begin{figure*}[!h]
    \centering
    \includegraphics[height=.4\textwidth,trim={0cm 0.0cm 0.0cm 0cm},clip]{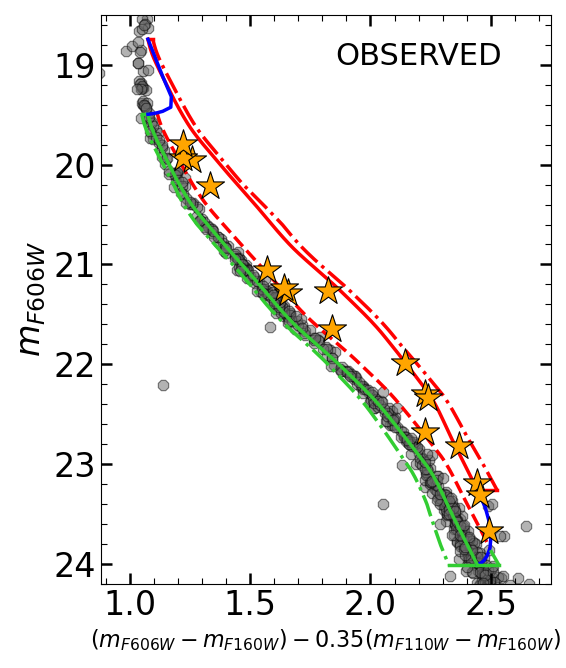}
    \includegraphics[height=.4\textwidth,trim={0cm 0.0cm 0.0cm 0cm},clip]{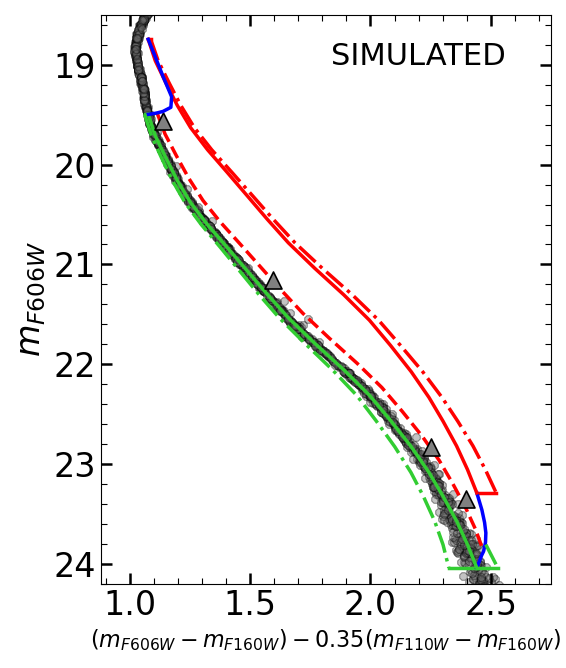}
     \caption{
     Procedure to estimate the fraction of binaries among lower MS stars with masses of $0.39-0.75\ M_{\odot}$, following the same methodology explained in section \ref{sec:3}. The colours and symbols are consistent with those in Fig.\,\ref{fig:Fig2}
     }
    \label{fig:Fig4}

\end{figure*}


To determine the incidence of binaries among 1P and 2P stars, we compared the observations with simulated diagrams hosting binaries  generated from ASs. The latter exhibit various combinations for the incidence of 1P binaries, 2P binaries, and mixed binary systems, with the fraction of each group of binaries ranging from 0 \% to 100\% in steps of 1\%.

We analyzed the simulated diagrams by using the same procedure adopted for the real stars. 

We derived the CDF of the binaries for each combination of the fraction of the total number of binaries in 1P-1P ($f^{\rm 1P}$), 2P-2P ($f^{\rm 2P}$), and 1P-2P ($f^{\rm mix}$) mixed binaries and compared it with the corresponding distribution derived for real stars by means of $\chi^2$. The simulation that provides the best match with the observation correspond to $f^{1P}=97^{+1}_{-3}\%$,$f^{2P}=3\pm 1\%$, and no mixed binaries $f^{\rm mix}=0^{+1}_{-0}\%$.

The uncertainties are derived through a Monte Carlo analysis using 1,000 simulated binary samples, each containing the same number of binaries as the observed sample and using the same the best-fit fractions for 1P, 2P, and mixed binaries. For each simulation, we calculated the values of $f^{\rm 1P}$, $f^{\rm 2P}$, and $f^{\rm mix}$ as we did for the real stars. The upper and lower uncertainties are represented by the $25^{th}$ and $75^{th}$ percentiles of the resulting values.

The left and middle panels of Fig.\,\ref{fig:Fig6} shows the $m_{F606W}$ vs.\,($m_{F606W}-m_{F160W}$)-0.35($m_{F110W}-m_{F160W}$) and $m_{F160W}$ vs.\,$m_{F110W}-m_{F160W}$ simulated diagrams, respectively, that provide the best match with the observations. 
In the right panels we compare the corresponding verticalized $m_{F160W}$ vs.\,$\Delta_{F110W,F160W}$ diagram (top) and the $\Delta_{F110W,F160W}$ CDF (bottom) for simulated and observed binaries.

The 1P and 2P binary fractions found in our analysis correspond to a binary incidence within the 1P $(f_{\rm 1P,bin})$ and 2P $(f_{\rm 2P,bin})$ populations equal to, respectively, $ f_{\rm 1P,bin}=6.41 \pm 1.74\%$  and $f_{\rm 2P,bin}=0.29 \pm 0.22\%$.

\section{Radial and mass distribution of the binaries in NGC\,288}
\label{radial}
The fact that the fiducial lines of 1P and 2P binaries are nearly coincident in the $m_{\rm F606W}$ vs.\,($m_{\rm F606W}-m_{\rm F160W}$)-0.35($m_{\rm F110W}-m_{\rm F160W}$) 
 plane  makes this pseudo-CMD and ideal tool to infer the overall binary fraction.
To do this, we used the method described in section \ref{sec:3}, which is illustrated in
 Fig.\,\ref{fig:Fig4}, where we compared the observed pseudo-CMD and the simulated pseudo-CMD for single stars.
 The results are summarized in Table\,\ref{tab:binary fraction} and show that the fraction of binaries with q$>$0.5 in the studied field of view is $3.93\pm0.95\%$.
  This value is consistent, within one sigma, with that inferred from the $m_{\rm F814W}$ vs.\,$m_{\rm F606W}-m_{\rm F814W}$ CMD.
 %
 %
\begin{figure}
    \centering
    \includegraphics[width=.45\textwidth,clip]{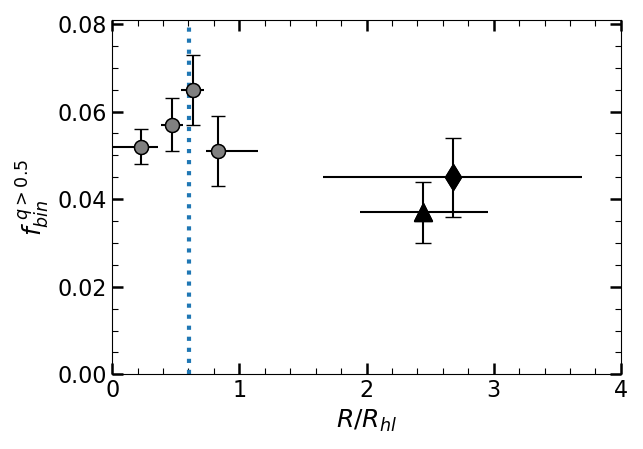}

    \caption{ 
    Fraction of binaries with $q>0.5$ as function of the radial distance from the center of NGC 288 (the radial distance R is normalized by the half-light radius, $R_{hl}$)  . Grey circles are measured by \cite{milone2012}, black diamond and triangle refer to the binary fraction measured in the magnitude range $19.0\leq m_{F814W}\leq23$ and $19.5\leq m_{F606W}\leq24$, respectively. Blue dotted line marks the core radius (1.35 arcmin). Horizontal and vertical black lines represent the radial coverage of each data point and the uncertainty associated to the measured fraction of binaries, respectively. 
    }
     \label{fig:Fig10}
\end{figure}

\cite{milone2012} studied the binaries within the half-light radius of NGC\,288 and derived a fraction of binaries with $q>0.5$ of $5.6 \pm 0.4 \%$, which is slightly larger than that observed in our work, at a distance from the center of the cluster of about three half-light radii. The radial behavior of the binary fraction is illustrated in Fig.\,\ref{fig:Fig10} and is consistent with a flat distribution with some hints of a larger binary fractions in the center. 
  
The $m_{\rm F606W}$ vs.\,($m_{\rm F606W}-m_{\rm F160W}$)-0.35($m_{\rm F110W}-m_{\rm F160W}$) diagram allows us to analyze the binaries  with  $19.5<m_{\rm F606W}<24$ mag 
which corresponds to a mass range between 0.39 and 0.75 $ M_{\odot}$. This mass range is slightly smaller than that studied with the $m_{\rm F814W}$ vs.\,$m_{\rm F606W}-m_{\rm F814W}$ CMD ($0.36-0.75\ M_{\odot}$). To investigate the relation between the binary fraction and the mass of the primary star, we divided the studied F606W magnitude interval into three bins of $1.5$ mag each and calculated the fraction of binaries with q$>$0.5 in each of them. As indicated in Table\,\ref{tab:binary fraction2}, the fractions of binaries in the three luminosity intervals are consistent with each other at one-sigma level. We conclude that there is no evidence of a dependence of the binary fraction on the stellar mass.

\label{binary incidence}

\section{Simulations}

To further investigate the dynamical processes influencing the observed profiles of NGC 288, we conducted a Monte Carlo simulation following the dynamical evolution of a multiple-population GC and its binary population. We would like to emphasize that this simulation is not intended to precisely model NGC\,288, but rather to provide insights into the dynamics of multiple-population GCs and their binary stars. 

\begin{figure}
    \centering
    \includegraphics[width=0.45\textwidth]{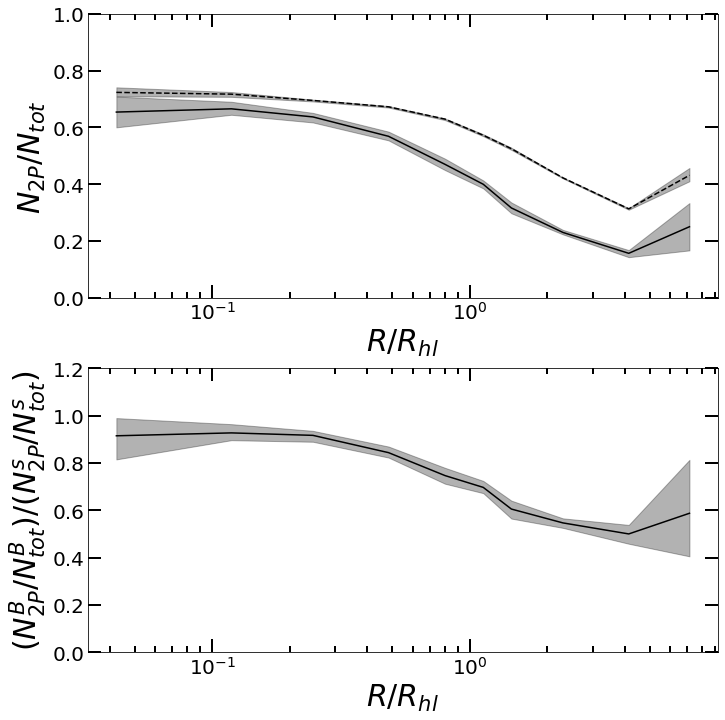}  
    \caption{\textit{Top.} Radial profile of the fraction of binaries (solid line) and single stars (dashed line) belonging to the 2P population (the projected distance from the cluster's center, $R$, is normalized by the half-light radius, $R_{hl}$). \textit{Bottom.} Radial profile of the ratio of the fraction of binaries to the fraction of single stars belonging to the 2P population. The shaded regions represent the 25th and 75th percentiles of the 100 realizations of the 2D random projection.}
    \label{fig:Figure A}
\end{figure}

\begin{figure}[!t]
    \centering
    \includegraphics[width=0.45\textwidth]{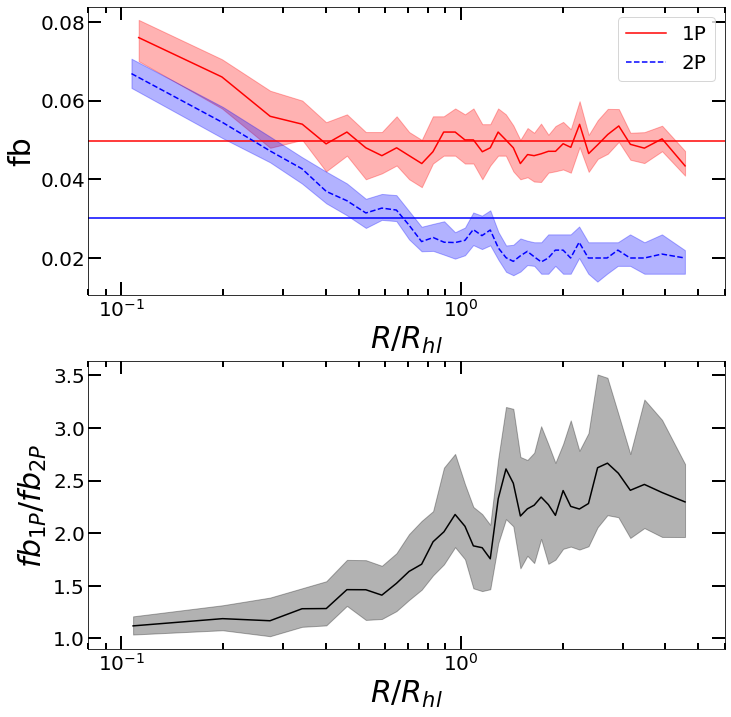}  
    \caption{\textit{Top}. The radial profile of the binary incidence in the 1P population (red line) and the 2P population (blue line). The global binary fractions for each population are indicated by the horizontal lines. \textit{Bottom}. The ratio of the binary incidence in the 1P population to that in the 2P population. The shaded regions represent the 25th and 75th percentiles of the 100 realizations of the 2D random projection.}
    \label{fig:Figure B}
\end{figure}

The simulation was performed using the \textsc{mocca} code on the Quartz supercomputer at Indiana University. The \textsc{mocca} code accounts for the effects of two-body relaxation, binary-single star and single-single star close encounters, stellar evolution, and tidal truncation (see, e.g., \citealt{Hypki&Giersz2013}, \citealt{Hypki+2024} for further details about the code). The model begins with $10^6$ stars with stellar masses following a \citet{Kroupa2001} initial mass function between 0.1 $M_{\odot}$ and 100 $M_{\odot}$ and a $10 \%$ binary fraction. The spatial distribution of the 2P and 1P populations follow King models \citep{King1966} with central dimensionless potential values, $W_0$, equal to 7 and 5, respectively. 
Following the general results of a number of simulations modeling the formation of the 2P population (see e.g. \citealt{dercole08,bekki10,calura19}), we start our simulation with the 2P population initially centrally concentrated in the inner regions of the 1P system. For the simulation presented here we adopted an initial ratio of the 2P to the 1P half-mass radius equal to 0.05 (see \citealt{vesperini21}, \citealt{livernois24} for other simulations  starting with this and other values of the 2P concentration; see also \citealt{hypki22,Hypki+2024} for a survey of models exploring various initial conditions).
The initial configuration placed $15\%$ of the total initial mass
of the cluster in the 2P population (the final fraction of the total mass in 2P stars at 12 Gyr is equal to about 0.51 and similar to the 2P fraction, $\sim 0.46 \pm 0.03$, found for NGC288 in \citealt{Milone+2017}). To align the simulation with the observational constraints, for the results shown in Figures \ref{fig:Figure A} and \ref{fig:Figure B}  we restricted the analysis to single stars on the MS with masses between 0.36 and 0.75 $M_\odot$, and binary stars to MS-MS binaries with mass ratios q >0.5, where the
primary star has a mass between 0.36 and 0.75 $M_\odot$.

The simulation was evaluated at an evolutionary age of 12 Gyr, consistent with the typical age of old GCs. At 12 Gyr, the fraction of the total mass in 2P stars in our model is about 50 percent, a value well within the range of those found in Galactic GCs (see e.g. \citealt{Milone+2017}). 

\begin{figure}
    \centering
    \includegraphics[width=0.35\textwidth]{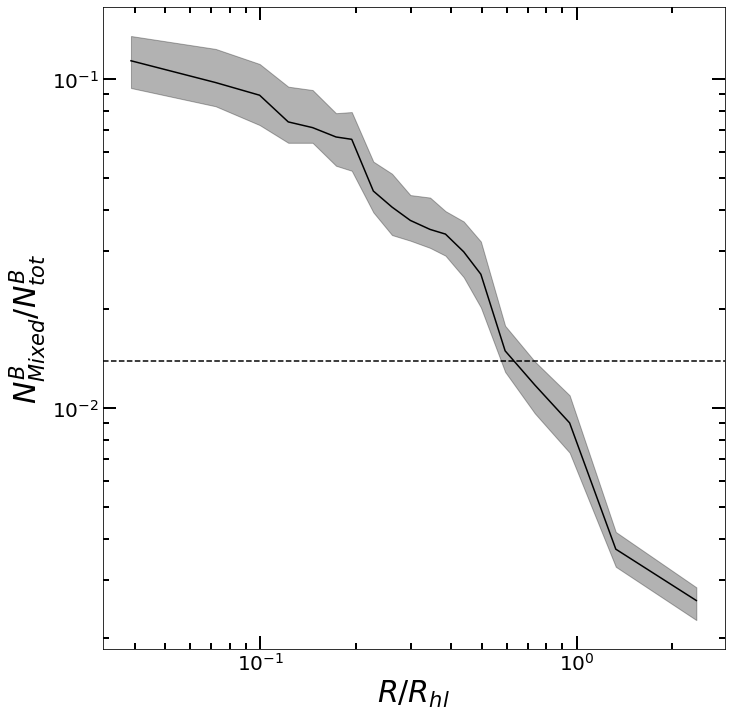}  
    \caption{The radial profile of the fraction of binaries in mixed binaries. The shaded
regions represent the 25th and 75th percentiles of the 100 realizations
of the 2D random projection. The global fraction of binaries in mixed binaries is indicated by the horizontal dashed line.}
    \label{fig:Figure C}
\end{figure}

Each radial profile presented in Figures \ref{fig:Figure A}, \ref{fig:Figure B}, and \ref{fig:Figure C} was derived from 100 different random realizations of the 2D projection of the 3D data generated by the simulation. For each plot, we display the median profile along with shaded regions representing the 25th and 75th percentiles of the 100 realizations.

One of the results emerging from the observational analysis in the previous sections is the differing degrees of 1P/2P spatial mixing between single stars and binary stars. As shown by \cite{Jang2022} and discussed later in Section \ref{summary}, 
 the  fraction of 2P single stars does not vary with the distance from the cluster's center, while the fraction of 2P binaries decreases with clustercentric distance. In the top panel of Fig.\,\ref{fig:Figure A} we present the radial profiles of the fraction of binaries (solid line) and single stars (dashed line) in the 2P population. This figure highlights a more significant degree of spatial mixing among single stars, reflected in a flatter radial profile, whereas the binary population exhibits a steeper radial gradient starting from smaller radii. This pattern is consistent with binary stars undergoing a slower mixing process compared to single stars. This behavior is consistent with the findings of \citet{Hong+2019}, who showed that binary stars experience additional dynamical interactions (such as ionization and ejection) that can delay their mixing compared to single stars.

In the lower panel of Fig.\,\ref{fig:Figure A}, we present the radial variation of the ratio between the fraction of binaries and the fraction of single stars in the 2P population. A steep radial gradient is observed, further emphasizing that binaries are less mixed than single stars. This result confirms that a GC can exhibit varying degrees of spatial mixing for single and binary stars, consistent with the observational findings for NGC 288.
The top panel of Fig.\,\ref{fig:Figure B} shows the radial profile of binary incidences in each population, while the bottom panel displays the ratio of 1P to 2P binary incidences. In the outer regions, the binary incidence in the 1P population is notably higher than that in the 2P population. However, as we move towards the inner regions, the ratio of the incidences tends to 1 (see also the discussion in \citealt{milone2020}). The radial variation observed in our simulation aligns with the trend seen in the observational data for NGC 288: \citet{milone2020} reported similar binary incidences for the 1P and 2P populations in the cluster's inner regions, while the results presented in this paper show that the outer regions are characterized by a higher binary incidence in the 1P population compared to the 2P population.

In the top panel of Fig.\,\ref{fig:Figure B}, we also present the global binary incidences for both populations, demonstrating that the 2P global binary incidence is smaller than that of the 1P population. This difference arises from the more centrally concentrated spatial distribution and the higher binary ionization rate in the 2P population (see \citealt{vesperini11}; \citealt{hong15,hong2016a} for the first studies of the dynamics of binaries in multiple-population clusters based on more idealized N-body simulations and  \citealt{sollima22,hypki22,Hypki+2024} for more recent and realistic Monte Carlo simulations). 

As 1P and 2P single and binary stars mix and interact, binary-binary and single-binary encounters can result in the formation of "mixed binaries", consisting of one 1P star and one 2P star. The encounter rate, which increases towards the cluster center, plays a key role in the formation of these mixed binaries. Fig.\,\ref{fig:Figure C} shows that the fraction of the total binary population in mixed binaries in our model is highest near the cluster center and declines steeply in the outer regions (given the small total number of mixed binaries, to better illustrate the radial variation of the mixed binary fraction in this figure we used all the binaries in the model at 12 Gyr and not just those following the observational selection criteria). The radial variation observed in our simulation aligns with the observational trend found in NGC 288 with the data analyzed in this paper and in \citet{milone2020}.

\label{sec:6}

\begin{figure}
    \centering
    \includegraphics[width=.45\textwidth,clip]{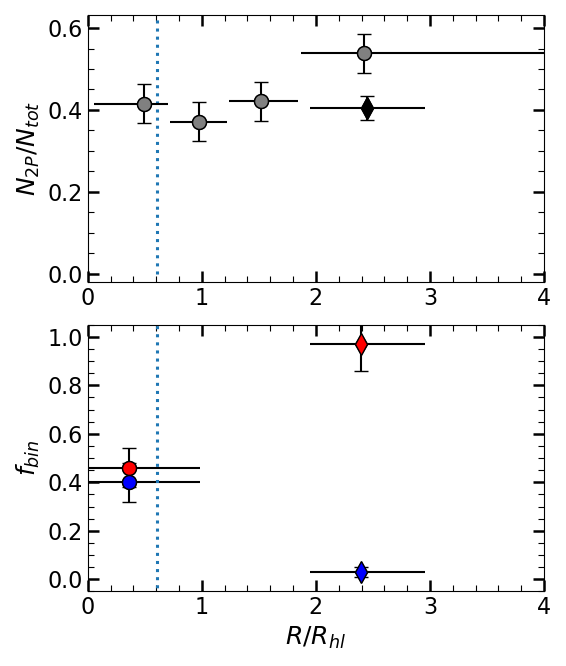}
    \caption{ 
     \emph{Top}. Fraction of 2P stars as function of the radial distance from the cluster center. Grey circles are measured by \cite{Jang2022} from ground-based photometry, black diamond are form this research. Horizontal and vertical black lines represent the radial coverage of each data point and the corresponding errorbars, respectively.
    \emph{Bottom}. Fraction of 1P and 2P binary with respect to the total number of binaries, $\rm f_{bin}$, as function of the radial distance from the cluster center. Circle markers are from \cite{milone2020} and diamond markers are from the present work. Red and blue colors refer to 1P and 2P binary fraction, respectively. 
    }
     \label{fig:Fig11}
\end{figure}

\section{Summary and discussion}
\label{summary}
Binary systems provide valuable insight into the formation and dynamical history of multiple stellar populations in GCs.  
Several formation scenarios suggest that 2P stars form in a high-density subsystem embedded in a less dense 1P cluster; the differences in the 1P and 2P environment set at the time  of the cluster formation can affect the survival and evolution of 1P and 2P binary stars.  Therefore, the present-day binary fraction of 1P and 2P stars are valuable indicators of the distinct environments in which these stellar populations formed and their subsequent evolution. 

Previous photometric studies have mainly concentrated on binaries within the innermost regions of GCs, within $\sim$1 $R_{hl}$ \citep{milone2020}, while spectroscopic studies tend to focus on more external regions \citep{dorazi2010a, lucatello2015a, dalessandro2018a}. To fully capture the possible differences in the incidence of 1P and 2P binaries predicted by theoretical studies (see \citealt{vesperini11, hong15,sollima22,hypki22,Hypki+2024}), along with the occurrence of mixed binaries (comprising one 1P and one 2P star) formed through interactions with other binaries and single stars, a comprehensive and consistent analysis of the radial variation in 1P and 2P binary frequencies from the cluster center to its outer regions is essential.

The central regions are highly influenced by the cluster’s dynamic evolution, where frequent interactions in these dense environments often erase traces of primordial binaries. Conversely, the outer, less dense regions are more effective at preserving this information.
 
In this study, we examined binaries in the outer region of NGC\,288 at $\sim 5.4\ \rm arcmin$ from the cluster center ($\sim 2.5\ R_{hl}$) and combine our findings with results obtained by \cite{milone2020} in the cluster center. 

To do this, we used HST images collected in the F606W and F814W bands with ACS/WFC, as well as in the F110W and F160W bands with WFC3/IR. 
To measure the binary fraction among 1P and 2P stars, we adopted the method developed by \cite{milone2020}, thus providing homogeneous investigation of binaries in both the cluster center and in the external field.

First, we disentangled the two stellar populations using the $m_{\rm F160W}$ versus $m_{\rm F110W}-m_{\rm F160W}$ CMD, which effectively disentangles stellar populations with different O abundance below the MS knee. 
 Next, we identify a new photometric diagram where the sequences of the two stellar populations overlap, the $m_{\rm F606W}$ versus  $(m_{\rm F606W}-m_{\rm F160W})-0.35\ (m_{\rm F110W}-m_{\rm F160W})$ pseudo-CMD, and we selected a sample of 12 binaries.
 To infer the incidence of 1P and 2P binaries, we compared the cumulative distribution function of the selected binaries in the verticalized diagram $m_{\rm F160W}$ versus $\Delta_{\rm F110W, F160W}$ with those derived from a grid of simulated diagrams with different fractions of 1P, 2P, and mixed binaries.




Our analysis revealed a predominance of 1P binaries compared to 2P binaries in the outer regions of NGC\,288 where $97^{+1}_{-3} \%$ of the binaries are 1P-1P binaries, $3\pm\ 1 \%$ 2P-2P binaries, and no mixed binaries. 
Assuming that 2P stars represent $40.5\%$ of the total stellar population (Fig.\,\ref{fig:Fig11}), we find that the fraction of 1P binaries relative to the total number of 1P stars is approximately 20 times larger than that of 2P stars. Considering a fraction of binaries with q$>$0.5 of $3.93\pm0.95\%$, we infer that the incidence of binaries is $6.4\pm\ 1.7 \%$ within the 1P population and $0.3\pm\ 0.2 \%$ for the 2P population.

In the cluster's center, instead, \cite{milone2020} measured a similar fraction of 1P and 2P binaries, respectively $46\pm 8\%$ and $40\pm 8\% $ (Fig.\,\ref{fig:Fig11}) and similar binary incidences within each population (with a ratio of the 1P to the 2P binary incidences equal to $1.0\pm0.3$).  The radial variation in the 1P and 2P binary incidences is in general agreement with the predictions of models of multiple-population clusters starting with a denser 2P system embedded in the central regions of a less concentrated 1P cluster: the initial differences between the 2P and 1P lead to a more rapid binary disruption for the 2P population during the cluster's dynamical evolution (see Section \ref{sec:6}; see also \citealt{hong15,hong2016a}).
Our study also reveals a  difference between the degree of spatial mixing of single 1P and 2P stars and that of 1P 
and 2P binaries; as shown in the upper panel of Fig. \ref{fig:Fig11}  the fraction of 2P stars (as measured by \citealt{Jang2022}) does not vary with the radial distance from the cluster's center indicating that the two populations are  mixed and lost memory of the initial differences between their spatial distribution. The fraction of the total number of binary stars in 2P binaries, on the other hand, is characterized by a significant radial variation and, as discussed above, decreases in the cluster's outer regions (bottom panel of Fig. \ref{fig:Fig11}).
The differences between the degree of mixing of single and binary stars is also in general agreement with the results of the models  presented in Section \ref{sec:6} where we have shown that spatial mixing of the two populations occurs on a longer timescale for binaries than for single stars (see  \citealt{Hong+2019} for the first discussion of the various dynamical aspects behind the delayed mixing of 1P and 2P binaries, and \citealt{dalessandro2018a} for the possible manifestation of this difference on the kinematics of 1P and 2P stars in NGC\,6362).
Finally, we point out that our analysis did not find evidence of mixed (1P-2P) binaries in the cluster's outer regions, while \cite{milone2020} found a fraction of mixed binaries equal to $14\pm7\%$  in the central regions.
Mixed binaries have been predicted (\citealt{hong15,hong2016a}) to form in binary-binary and binary-single interactions during which one of the binary components is replaced by a star of a different population. As shown in the simulations presented in section \ref{sec:6}, these binaries are expected to form mainly in the cluster's inner regions where binary interactions are more frequent and the difference between the inner and outer fraction of mixed binaries is consistent with the predictions of our simulations.


\begin{acknowledgements}
This work has received funding from 
"PRIN 2022 2022MMEB9W - {\it Understanding the formation of globular clusters with their multiple stellar generations}" (PI Anna F.\,Marino), from INAF Research GTO-Grant Normal RSN2-1.05.12.05.10 -  (ref. Anna F. Marino) of the "Bando INAF per il Finanziamento della Ricerca Fondamentale 2022", and from the European Union’s Horizon 2020 research and innovation programme under the Marie Skłodowska-Curie Grant Agreement No. 101034319 and from the European Union – NextGenerationEU (beneficiary: T. Ziliotto). EV acknowledges support from NSF grant AST-2009193.
\end{acknowledgements}

\bibliographystyle{aa}
\bibliography{aanda}
\end{document}